\documentclass[aps,prc,twocolumn,superscriptaddress,showpacs,floatfix]{revtex4}
\usepackage{graphicx,amsmath,amssymb,bm}

\newcommand{\be}{\begin{equation}}
\newcommand{\ee}{\end{equation}}
\newcommand{\vlk}{V_{{\rm low}\,k}}

\newcommand{\fmi}{\, {\rm fm}^{-1}}
\newcommand{\mev}{\, {\rm MeV}}
\newcommand{\la}{\Lambda}
\newcommand{\oab}{O_{AB}}
\newcommand{\hw}{\hbar \omega}
\newcommand{\cspm}{$cs \pm 1\ $}

\begin{document}

\title{Shell-model phenomenology of low-momentum interactions}
\author{Achim Schwenk}
\email[E-mail:~]{schwenk@indiana.edu}
\affiliation{Nuclear Theory Center, Indiana University,
2401 Milo B. Sampson Ln, Bloomington, IN 47408}
\author{Andr\'es P. Zuker}
\email[E-mail:~]{Andres.Zuker@IReS.in2p3.fr}
\affiliation{Institut de Recherches Subatomiques, IN2P3-CNRS,
Universit\'e Louis Pasteur, F-67037 Strasbourg}

\begin{abstract}
  The first detailed comparison of the low-momentum interaction $\vlk$
  with $G$ matrices is presented. We use overlaps to measure
  quantitatively the similarity of shell-model matrix elements for
  different cutoffs and oscillator frequencies. Over a wide range, all
  sets of $\vlk$ matrix elements can be approximately obtained from a
  universal set by a simple scaling. In an oscillator mean-field
  approach, $\vlk$ reproduces satisfactorily many features of the
  single-particle and single-hole spectra on closed-shell nuclei, in
  particular through remarkably good splittings between spin-orbit
  partners on top of harmonic oscillator closures.
  The main deficiencies of pure two-nucleon interactions are
  associated with binding energies and with the failure to ensure
  magicity for the extruder-intruder closures. Here, calculations
  including three-nucleon interactions are most needed. $\vlk$ makes
  it possible to define directly a meaningful unperturbed monopole
  Hamiltonian, for which the inclusion of three-nucleon forces is tractable.
\end{abstract}


\maketitle

Microscopic nuclear structure studies fall in three categories.  For
local interactions, the Green's Function Monte Carlo (GFMC)
method~\cite{Pieper.Wiringa:2001,Pieper.Wiringa.Carlson:2004} leads to
exact solutions of the many-body Schr\"odinger equation by evaluation
of multi-dimensional integrals in coordinate space. The No-Core
Shell-Model (NCSM)~\cite{Navratil.Vary.Barrett:2000,Navratil.Caurier:2004}
relies on matrix diagonalizations in a harmonic oscillator
basis of $N \hw$ excitations with respect to a minimal $0 \hw$ space.
Convergence with $N \hw$ is slow for conventional nucleon-nucleon (NN)
interactions, which are replaced by effective interactions that are
model-space dependent. Both GFMC and converged NCSM methods are
limited at present to mass number $A \lesssim 12$.  The standard
Shell-Model (SM)~\cite{Caurier.Martinez-Pinedo.ea:2003} restricts
diagonalizations to $0 \hw$ spaces and treats higher excitations in
perturbation theory. It bypasses saturation problems by using a $G$
matrix~\cite{Hjorth-Jensen.Kuo.Osnes:1995} calculated at approximately
the experimental nuclear radius ($\hw \approx 40 A^{-1/3}$) and uses
experimental single-particle energies. Presently, exact SM
diagonalizations are possible for all semi-magic nuclei, and for
$A<70$ in full $0\hw$ spaces.

It has been traditionally assumed that NN interactions are strongly
repulsive at short distances, and therefore require resummations to
obtain ``pseudopotentials'' in a given model space. For fifty years 
the standard in nuclear physics has been the Brueckner-Bethe-Goldstone 
$G$ matrix, which is calculated from a NN potential $V$ by summing
two-particle ladders outside the model space,
\begin{equation}
\label{eq:G}
G_{ijkl}=V_{ijkl}-\sum_{\alpha\beta}\frac{V_{ij\alpha\beta} \, 
G_{\alpha\beta kl}}{\epsilon_{\alpha}+\epsilon_{\beta}
-\epsilon_i-\epsilon_j+\omega_s} \,,
\end{equation}
where $\epsilon_x$ are unperturbed (usually kinetic) energies; $ij$ 
and $kl$ denote orbits in the model space and $\alpha \beta$ orbits 
outside it, while the starting energy $\omega_s$ is treated as a
free parameter. Interestingly, the $G$ matrix approach yields SM
interactions that are, up to an overall multiplicative factor of
all matrix elements, roughly independent of the NN potential and
the starting energy used~\cite{Abzouzi.Caurier.Zuker:1991,
Dufour.Zuker:1996,Caurier.Martinez-Pinedo.ea:2003}. The shortcoming
of the $G$ matrix is the ill-defined relationship between the
starting energy and the model space, thus precluding ab initio
calculations. 

An alternative to the $G$ matrix approach starts by noting that 
conventional NN interactions are
well-constrained by two-nucleon scattering data only for laboratory
energies $E_{\rm lab} \lesssim 350 \, {\rm MeV}$. As a consequence,
details of nuclear forces are not resolved for relative momenta $k >
2.0 \, {\rm fm}^{-1}$. Starting from a NN potential, the high-momentum
modes can be integrated out in free space using the renormalization
group. The resulting low-momentum interaction, called $\vlk$, only has
momentum components below a cutoff $\la$ and evolves with it so that
all low-energy two-body observables (phase shifts and deuteron binding
energy) are preserved.  For $\la \lesssim 2.0 \fmi$, all NN potentials
that fit the scattering data and include the same long-distance pion
physics collapse to a universal $\vlk$~\cite{Bogner.Kuo.Schwenk:2003}. 
$\vlk$ defines a new NN interaction without a strong core, that can
be directly used in nuclear structure calculations and therefore
eliminates all pseudopotential approximations.

\begin{figure*}[t]
\begin{center}
\includegraphics[scale=0.4,clip=]{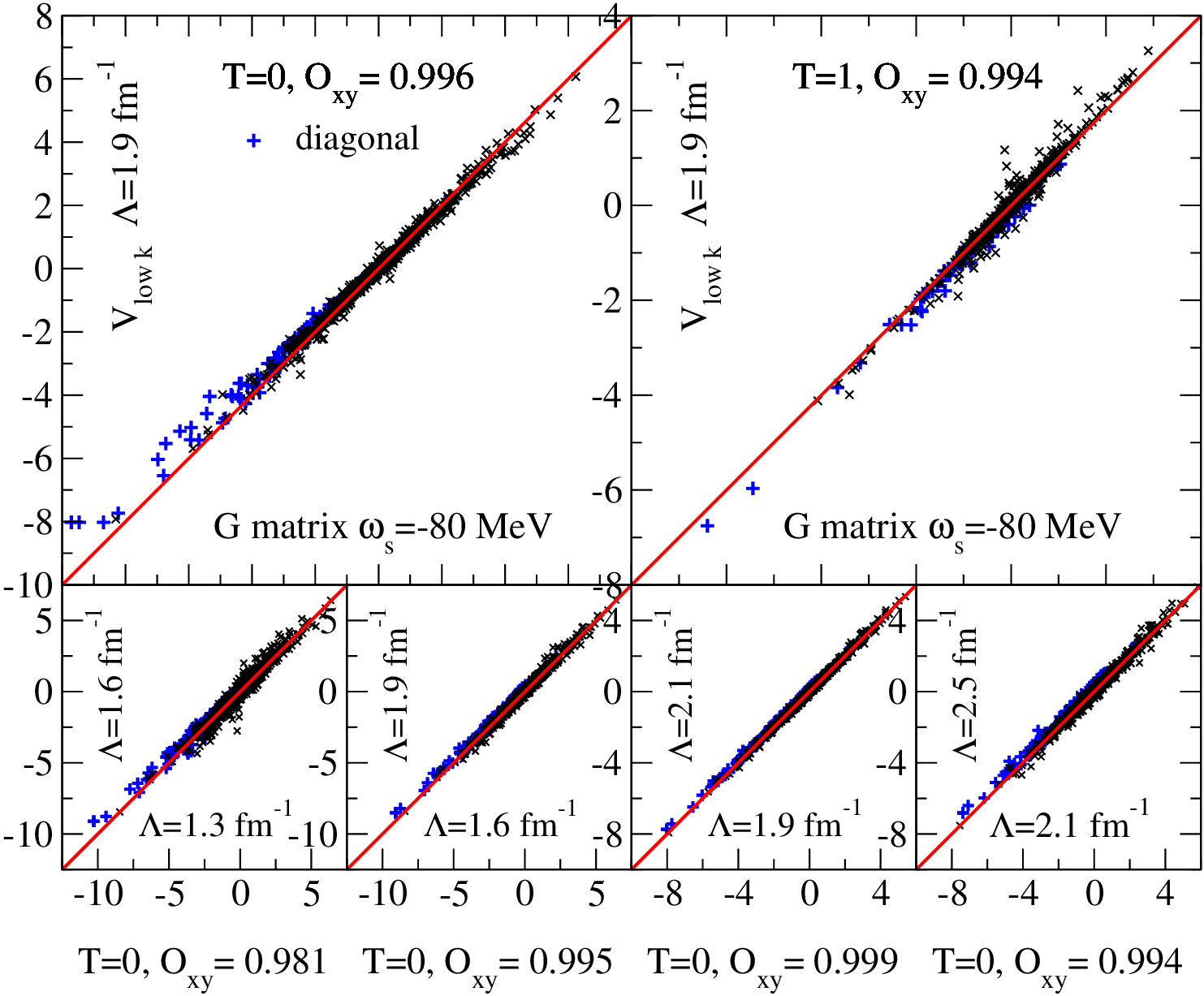}
\end{center}
\caption{\label{fig:corrplots} (Color online) Top: Correlation plots
  between $\vlk$ and $G$ matrix elements in a restricted space of 4 major
  shells. The matrix elements $V_{rstu}^{JT}$ are in ${\rm MeV}$ for
  $\hw = 14 \mev$ and we have highlighted the diagonal elements.
  $\vlk$ is derived from the Argonne $v_{18}$ potential and the $G$
  matrix is for Idaho A, computed with a rectangular Pauli operator 
  for 4 major shells and starting energy $\omega_s= - 80 \, 
  {\rm MeV}$~\cite{Dean.HjorthJensen:2004}. 
  Bottom: Correlation plots between $\vlk$
  matrix elements for different cutoffs. To facilitate the comparison,
  we have rescaled the $y$-axis set of $\vlk$ matrix elements by
  $\sigma_{\la_x}/\sigma_{\la_y}$ according to the approximate scaling
  law Eq.~(\ref{eq:uni}). The $\sigma$-ratios are from left to right:
  $1.003$, $1.061$, $1.050$ and $1.111$. We find even fewer scatters
  for the $T=1$ matrix elements. All $\oab$ given are $O_{AB}^T$.}
\end{figure*}

When only NN interactions are used, all microscopic approaches have a
common problem, related to poor binding and shell formation
properties. It reflects in a deteriorating agreement with experiment
as the number of particles increases (active particles for the $0
\hbar \omega$ SM). This leads to the conclusion that three-nucleon
(3N) interactions are necessary. In the case of $\vlk$, it has been
shown that chiral 3N forces can be adjusted to remove the cutoff
dependence and give perturbative contributions for $\la \lesssim 2.0
\, {\rm fm}^{-1}$ in light nuclei~\cite{Nogga.Bogner.Schwenk:2004}. In
addition, these low-momentum 3N forces drive saturation in 
nuclear matter~\cite{Bogner.Schwenk.ea:2004}.

In the first part of this paper, we compare $\vlk$ with $G$ matrices.
By studying the cutoff and oscillator frequency dependence of $\vlk$
matrix elements, we demonstrate a universal behavior. A similar
behavior exists for $G$ over a reasonable range of starting energies.
The second part is devoted to extracting the $\vlk$
monopole Hamiltonian, which is then used to calculate binding
energies, as well as single-particle and single-hole spectra on
closed-shell nuclei (\cspm spectra). Many features will turn out to be
in good agreement with data, and the discrepancies identify what is
expected of 3N forces and how they are crucial in heavier systems.
Our work shows that the mean-field produced by $\vlk$ is a valuable 
first approximation that will greatly simplify further perturbative 
or coupled cluster treatment, and the inclusion of 3N forces.
 
In Fig.~\ref{fig:corrplots}, we compare $\vlk$ to $G$ matrix elements
in 4 major shells. We find that both $T=0$ and $T=1$ matrix elements
are very similar. For a quantitative comparison, we define the
overlaps of interactions $A$ and 
$B$~\cite{Dufour.Zuker:1996,Caurier.Martinez-Pinedo.ea:2003}
\begin{gather}
\sigma_{AB}^2 = d_2^{-1} \sum_{rstu\Gamma} [\Gamma] \, W^{\Gamma}_{rstuA} \, 
W^{\Gamma}_{rstuB} \, ,
\label{eq:overlap}
\end{gather}
where $W_{rstu}^{JT} = V_{rstu}^{JT} - \delta_{rt} \, \delta_{su} W$ and
$d_2$ is the dimensionality of the two-particle space, each state being 
counted $[\Gamma]=(2J+1)(2T+1)$ times. Here, the interaction $V$ is referred 
to its centroid $W$, defined by $\sum_{rs\Gamma} [\Gamma] 
\, W^{\Gamma}_{rsrs}=0$. We also introduce normalized overlaps
\begin{gather}
{O}_{AB} = \frac{\sigma^2_{AB}}{\sigma_A \, \sigma_B} \, ,
\label{eq:normoverlap}
\end{gather}
with $\sigma_A=\sigma_{AA}$ and similarly $O_{AB}^T$ for matrix elements 
with the same $T$. Interactions that differ at most by a factor 
$\sigma_{A}/\sigma_{B}$ have $\oab = 1$. The overlaps between $\vlk$ and 
the $G$ matrix are $\oab > 0.99$. 

\begin{table}[t]
\begin{center}
\begin{ruledtabular}
\begin{tabular}{c|cccccc}
$W$ & -1.374 & -1.035 & -0.802 & -0.620 & -0.546 & -0.463 \\ \hline
$\sigma_A$ & 3.288 & 2.488 & 1.931 & 1.500 & 1.323 & 1.127 \\ \hline\hline
$\hw$ & 18.4 & 13.9 & 11.0 & 8.8 & 7.9 & 6.9 \\ \hline
18.4 & 1.000 & 0.992 & 0.978 & 0.961 & 0.952 & 0.941 \\
13.9 & 0.992 & 1.000 & 0.996 & 0.987 & 0.982 & 0.975 \\
11.0 & 0.978 & 0.996 & 1.000 & 0.997 & 0.995 & 0.990 \\
8.8 & 0.961 & 0.987 & 0.997 & 1.000 & 0.999 & 0.998 \\
7.9 & 0.952 & 0.982 & 0.995 & 0.999 & 1.000 & 0.999 \\
6.9 & 0.941 & 0.975 & 0.990 & 0.998 & 0.999 & 1.000 \\
\end{tabular}
\end{ruledtabular}
\end{center}
\caption{\label{tab:hwdep}
Centroids $W$, widths $\sigma_{A}$ and overlaps $\oab$ between
$\vlk$ matrix elements in 4 major shells for different $\hw$. 
$\vlk$ is derived from the Argonne $v_{18}$ potential for $\la=1.9 \fmi$. 
The values of $\hw$ correspond approximately from left to right to the 
double-magic nuclei at $A=4$, $16$, $40$, $90$, $132$ and
$208$. $G$ matrices follow the same behavior.}
\end{table}

Next, we compare $\vlk$ matrix elements for different cutoffs. Over
the studied range $\la=1.3 \ldots 3.0 \fmi$, we find again very large
$\oab$ overlaps, as shown in the bottom panels of Fig.~\ref{fig:corrplots}.
To facilitate the comparison, we have rescaled the $y$-axis set of
$\vlk$ matrix elements by the widths $\sigma_{\la_x}/\sigma_{\la_y}$.
Up to this overall factor, we find that $\vlk$ matrix elements are 
approximately cutoff-independent. From Table~\ref{tab:hwdep}, we 
find a similar behavior for sets with different $\hw$ at fixed cutoff.
These observations can be combined in an approximate scaling law
\begin{gather}
\vlk^{\la_1, \hw_1} \approx \frac{\sigma_{\la_1, \hw_1}}{\sigma_{\la_2, \hw_2}}
\, \vlk^{\la_2, \hw_2} \: \Rightarrow \:
\vlk^{\la, \hw} \approx \sigma_{\la, \hw} \, U \, ,
\label{eq:uni}
\end{gather}
where $U_{rstu}^{JT}$ is a set of two-body matrix elements,
approximately independent of $\la$ and $\hbar \omega$. The decrease of
the overlaps as the range of $\la$ or $\hw$ increases indicates that
different parts of the interaction may scale differently. However,
over a fairly wide range, $\sigma_{\la, \hw}$ follows a simple scaling
law,
\begin{gather}
\frac{\sigma_{\la, \hw_1}} {\sigma_{\la, \hw_2}} \approx 
\biggl( \frac{\hw_1}{\hw_2}\biggr)^{\alpha} \qquad 
\frac{\sigma_{\la_1, \hw}} {\sigma_{\la_2, \hw}} \approx 
\biggl(\frac{\la_2}{\la_1}\biggr)^{\beta} \, .
\label{eq:scaling}
\end{gather}
Empirically, we have found that the majority of matrix elements scale
with $\alpha=1$ and $\beta=1/2$.  Note that there can be cases when
$\oab$ is very large, but the scaling law is not simple. This happens
for instance for different starting energies in the $G$ matrices
computed from the Bonn potential for $\omega_s = -5 \ldots -140 \mev$
at fixed $\hw=40\,\mev$~\cite{HjorthJensen:1998}, $\oab \gtrsim 0.99$
and $\sigma$ ranges from 2.22 to 1.85.

The preceeding observations suggest that an overall multiplicative
factor approximately captures 
the evolution of the interaction with $\la$ or $\hw$. 
To this list, one would like to add $N$, the number of
$\hw$ excitations allowed in a model space. As shell model
calculations with any microscopic interaction demand very large $N$ to
converge to the exact result, it is imperative to define effective
interactions for spaces of lower $N\hw$ (usually $0\hw$ except for the
lightest nuclei). However, for $\vlk$ at $\la \approx 2 \fmi$,
the particle-particle channel becomes 
perturbative~\cite{Bogner.Schwenk.ea:2004}, the scaling
laws in Eq.~(\ref{eq:scaling}) start operating and the $0\hw$ spaces
provide a meaningful first approximation.  Therefore, we
shall proceed by using the bare $\vlk$ for some very simple $0\hw$
calculations. The question of $N$-scaling will be studied in future 
work, while
keeping in mind the evidence from $G$ matrices that perturbative
renormalizations amount to multiplicative factors, though different
parts of the interaction may scale
differently~\cite{Abzouzi.Caurier.Zuker:1991,Dufour.Zuker:1996}.

We start by studying binding energies of closed shell nuclei and \cspm
spectra. Unless otherwise noted, in what follows we use $\vlk$ derived
from the Argonne $v_{18}$ potential for $\la=1.9 \fmi$.  We have
checked that, for small cutoffs, the results are practically
independent of the precision nuclear force used for $\vlk$.

As a first approximation the \cspm states are single
determinants that do not involve configuration mixing and are
described by the monopole Hamiltonian $H_m$. In the oscillator basis,
$H_m$ contains a diagonal and a non-diagonal part. The latter is
needed for a Hartree-Fock calculation and produces further
correlations~\cite{Caurier.Martinez-Pinedo.ea:2003} that will be
neglected here.

In neutron-proton formalism, the diagonal monopole Hamiltonian has a 
kinetic and a potential part, $H^d_m=K^d+V_m^d$:
\begin{gather}
H_m^d =K^d+\frac{1}{2} \sum_{r_x,s_y} V_{r_xs_y} m_{r_x}
(m_{s_y}-\delta_{r_xs_y}\delta_{xy}) \, ,
\label{eq:vm}
\end{gather} 
where $x,y=n$ or $p$, and $m_{r_x}$ is the number of particles in
orbit $r$ for fluid $x$. The centroids $V_{r_xs_x}$ are defined 
in~\cite{Caurier.Martinez-Pinedo.ea:2003}. For an introduction to
monopole effects see~\cite{Zuker:2004}.

\begin{figure}[btp]
\begin{center} 
\includegraphics[scale=0.375]{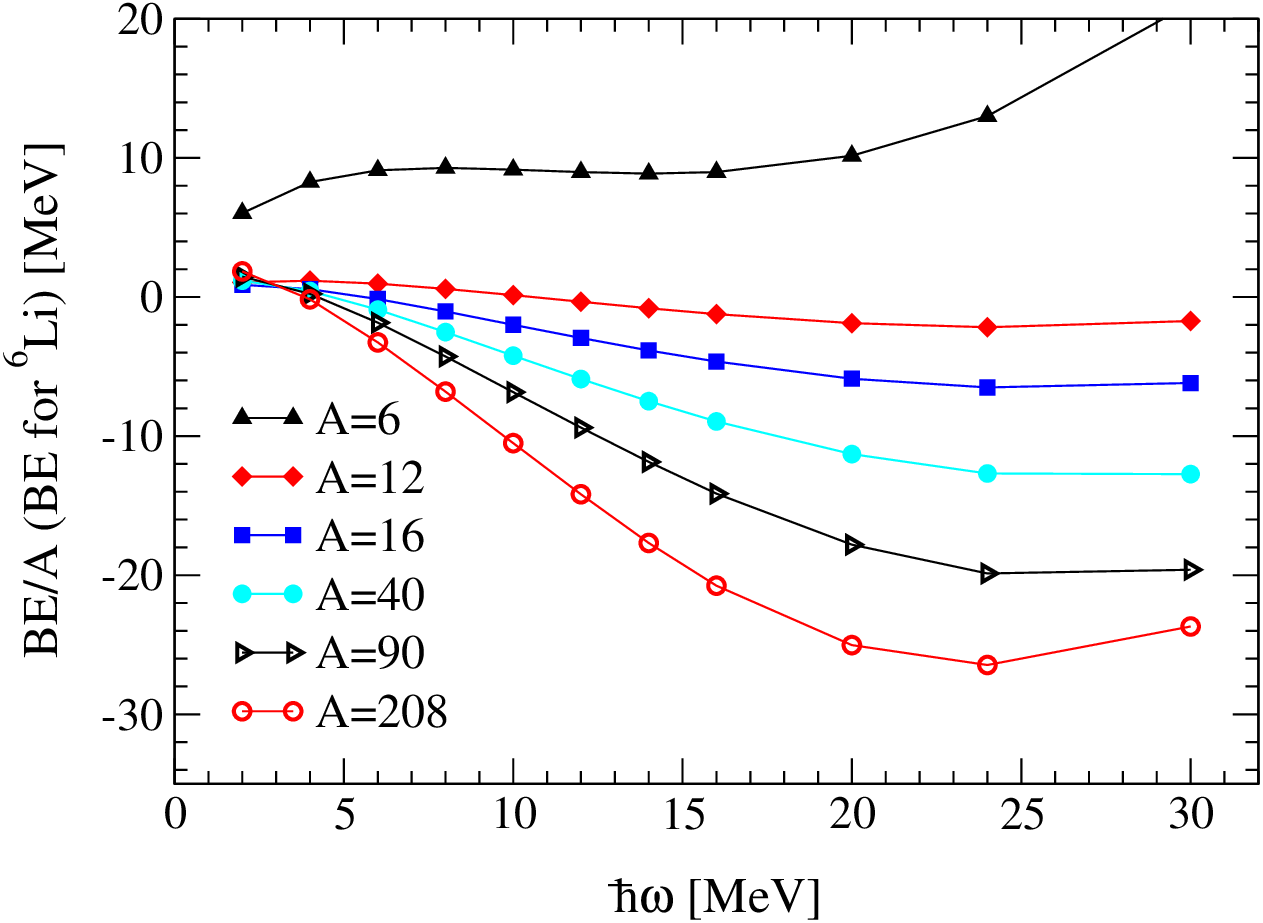}
\end{center}
\caption{\label{fig:be} (Color online) Results for the (negative)
  binding energies ($\text{BE} = \langle H_m^d \rangle$) obtained from
  Eq.~(\ref{eq:vm}) (Coulomb included schematically) by filling lowest
  oscillator orbits.}
\end{figure}

In Fig.~\ref{fig:be}, we show  the binding energies of closed shell 
nuclei. For small $\hw$, 
the system is dilute and the interaction behaves as a contact
($\delta$) force, leading to $\alpha=3/2$ in Eq.~(\ref{eq:scaling}).
We then observe a linear dependence on $\hw$, $\alpha = 1$, over the
range in which the saturation minimum should occur once correlations
and 3N forces are included. For large $\hw$, the interaction
effectively becomes long-ranged ($\alpha$ decreases) and the kinetic
energy takes over. The binding energies are cutoff dependent without
3N forces.  For example, in $^{40}$Ca at $\hw=12 \mev$, we have
$BE/A=8.24$, $5.89$ and $4.52 \mev$ for $\la=1.6$, $1.9$
(Fig.~\ref{fig:be}) and $2.1 \fmi$ respectively.  When higher
excitations are allowed, the $\hw$ dependence becomes weaker (as seen
for example in~\cite{Coraggio.Itaco.ea:2003}), and an exact result
will be independent of $\hw$.  Fig.~\ref{fig:be} indicates that, for
medium-mass and heavy nuclei, 3N interactions should provide a strong
repulsion at large $\hw$.

The case of $^{6}$Li suggests a special behavior in the light nuclei
in that a minimum is achieved at a reasonable $\hw = 14 \mev$,
although the total energy is still positive due to the
kinetic-potential competition: $\langle K^d \rangle = 88.0 \mev$ and
$\langle V^d_m \rangle = - 81.3 \mev$. Therefore, configuration mixing
and 3N contributions should lead to adding $40 \mev$ to $\langle V^d_m
\rangle$, a plausible
expectation~\cite{Navratil.Caurier:2004,Coraggio.Itaco.ea:2003}.
Note that a reduction of $\vlk$ to the $0\hw$ space would demand a
factor $\approx 1.4$ increase in $\langle V^d_m \rangle$ to achieve
the correct result: Again a plausible expectation, suggesting that for
medium-mass and heavy nuclei, the 3N interactions would have to
provide an even larger repulsion at large $\hw$.
\begin{figure}[btp]
\begin{center}
\includegraphics[scale=0.375,clip=]{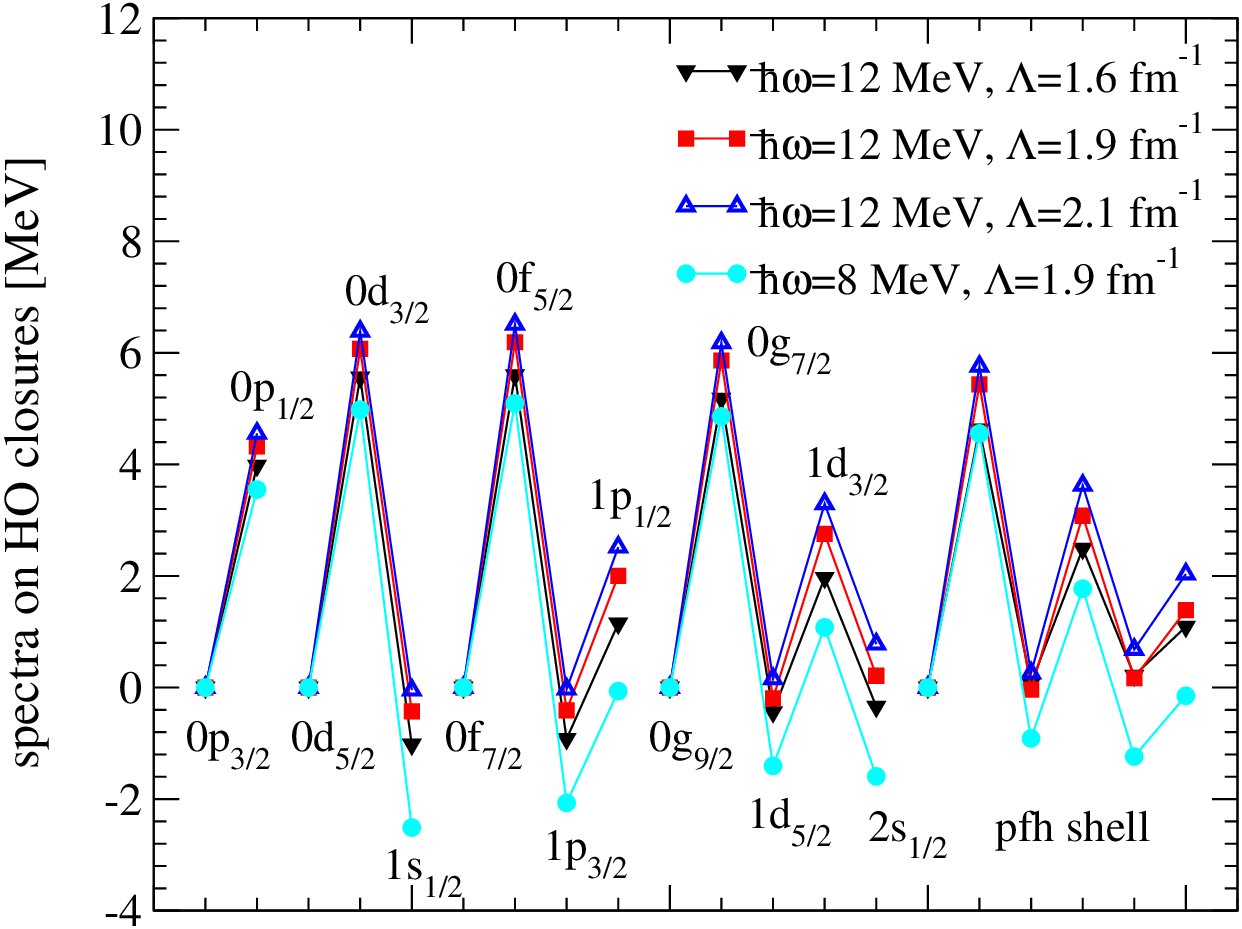}
\end{center}
\caption{\label{fig:eho} (Color online) Single-particle spectra on
  harmonic oscillator closures. Lines (meant to guide the eye) join
  points belonging to the same major shell. All energies are measured
  from the largest $j$ subshell in each major shell. We have used
  Eq.~(\ref{eq:scaling}) to refer all $\vlk$ matrix elements to
  $\la=1.9 \fmi$ and rescaled by
  $(\hw/\hw_0)^2=\bigl((p+2)^2[(p_0+2)^3+3]/(p_0+2)^2[(p+2)^3+3]^2\bigr)^2$
  (see text), which corresponds to the mass number $A$ appropriate to
  each major shell with principal quantum number $p+1$. $p_0=2\, (4)$
  correspond to the $pf\, (pfh)$ shell and $\hw_0=12\, (8) \mev$.}
\end{figure}

To obtain further insight, we study the \cspm spectra. One known
problem of an NN-only description is the failure to ensure the $N,Z
= 28, 50, 82, \ldots$ extruder-intruder
magicity~\cite{Pasquini.Zuker:1978,Caurier.Martinez-Pinedo.ea:2003}.
Our results will suggest that this is due to two basic shortcomings, 
the ``$\bm{l} \cdot \bm{l}$'' and ``$\bm{j_{max}}$'' anomalies,
that 3N interactions are expected to remedy. These are small effects 
$\sim A^{1/3}$ compared to the binding energies.

First, we show the single-particle spectra on top of harmonic oscillator 
closures in Fig.~\ref{fig:eho}. We have scaled all matrix elements to the 
$\hw$ corresponding to each major shell. For this we use $\alpha=2$ 
in Eq.~(\ref{eq:scaling}) to comply with the physical imperative that 
splittings should go asymptotically as $A^{-1/3}$, and we use
$\hw =35.59 A^{1/3}/\langle r^2
\rangle \mev$, with experimental mean-square radius $\langle r^2
\rangle = 0.943 \, A^{2/3} \, (1+2/A) \, 
{\rm  fm}^2$~\cite{Duflo.Zuker:2002} and $A \approx 2(p+2)^3/3$, where $p$
is the principal oscillator number at the Fermi level. Our results for
three values of the cutoff (using the empirical $\beta=1/2$) and
two values of $\hw$ are shown in Fig.~\ref{fig:eho}. If the
scaling laws were perfect, the four patterns should collapse into one.
They closely do for different cutoffs but less so for different
$\hw$, indicating that the constant $\alpha=2$ is too crude.

The splittings between spin-orbit partners in Fig.~\ref{fig:eho} agree
well with experiment, e.g., about $6.0 \mev$ for both the $d$ orbits
in $^{17}$O and the $f$ orbits in $^{41}$Ca. The $sdg$ and $pfh$
spectra on top of oscillator closures are not directly available but
the necessary information can be extracted from the known \cspm
spectra up to the Pb region. Our results correspond nicely
to those determined empirically~\cite{Duflo.Zuker:1999}, e.g., $4.6
\mev$ for the $h$ orbits as in Fig.~\ref{fig:eho} for $\hw=8 \mev$,
$\la=1.9 \fmi$.  However, what works well in~\cite{Duflo.Zuker:1999}
is a combination of one-body $\bm{l} \cdot \bm{s}$ and $\bm{l} \cdot
\bm{l}$ terms. The latter is attractive beyond the $pf$ shell, thus
favoring high $l$ orbits. Fig.~\ref{fig:eho} shows the opposite: low
$l$ orbits are always depressed with respect to the $\bm{l} \cdot
\bm{s}$ standard, and the effect grows bigger in heavier nuclei. This
is the $\bm{l} \cdot \bm{l}$ anomaly.

\begin{figure}[btp]
\begin{center}
\includegraphics[scale=0.35,clip=]{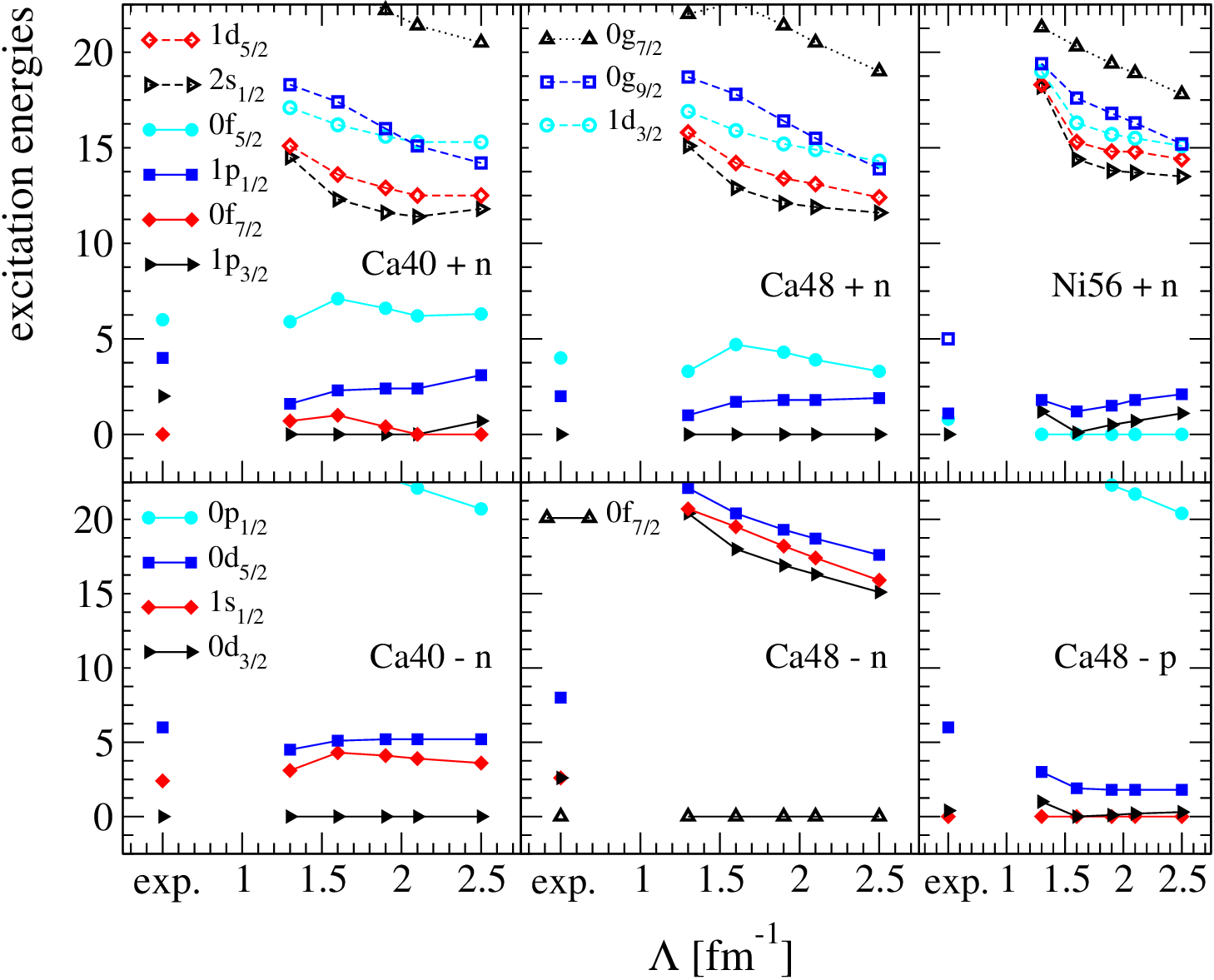}
\end{center}
\caption{\label{fig:cspm1} (Color online) \cspm spectra in the $pf$
  region for a wide cutoff range. The experimental $0g_{9/2}$ energy
  in $^{57}$Ni is an estimate.}
\end{figure}

Splittings for hole states on harmonic oscillator closures are known
in $^{15}$O and $^{39}$Ca ($\approx 6 \mev$ in both). For $^{15}$O we
obtain about half the observed value, a mean-field result similar
to~\cite{Pieper.Pandharipande:1993}.  (Note that this hole splitting
is not the same as the particle one in Fig.~\ref{fig:eho}). For
$^{39}$Ca we are close at $5 \mev$, as can be seen in the bottom-left
panel of Fig.~\ref{fig:cspm1}.

This figure compares the \cspm spectra in the $pf$ shell with
available data~\cite{NNDC} as summarized in~\cite{Duflo.Zuker:1999},
where it is shown that a six-parameter monopole Hamiltonian can
describe equally well \cspm spectra for $A>60$ and $A<60$. The latter
are mostly represented by the $pf$ region, which therefore exhibits
the basic mechanisms of shell formation (see
also~\cite{Caurier.Martinez-Pinedo.ea:2003,Zuker:2004}). This
one-plus-two-body phenomenology reveals what can be explained
well by realistic NN potentials, and what cannot must therefore
be ascribed to other (presumably 3N) mechanisms. Our task is to
establish the distinction.


The results for low-lying $pf$ levels in the upper panel of
Fig.~\ref{fig:cspm1} are approximately cutoff-independent and the
minor discrepancies are cured by SM calculations, which push up the
$1p_{3/2}$ orbit in $^{41}$Ca, do not change the good spectrum in
$^{49}$Ca, and also push up by $1 \mev$ the $0f_{5/2}$ orbit in
$^{57}$Ni~\cite{Caurier.Martinez-Pinedo.ea:2003,Nowacki:2002}.  For
the high-lying $sdg$ shell levels, the splittings between spin-orbit
partners agree with the empirical values, but the $\bm{l}\cdot \bm{l}$
anomaly is much stronger than in Fig.~\ref{fig:eho}, and partly
responsible for the $\sim 10 \mev$ underbinding of the $0g_{9/2}$
particle orbit in $^{57}$Ni.  A similar discrepancy shows for the
$0f_{7/2}$ hole orbit in $^{47}$Ca, underbound by $\sim 10 \mev$ with
respect to its $sd$ partners. Such shortcomings are responsible for
the failure of NN-only interactions to ensure the $N,Z=28, 50, 82
\ldots$ extruder-intruder closures.  This can be directly checked
through the standard measures of magicity, the gaps defined as
$g(cs,x)=2BE(cs)-BE(cs+x)-BE(cs-x)$ ($x=n$ or $p$). The ground state
spins are those of Fig.~\ref{fig:cspm1}. Rounded calculated
(experimental) gaps are (in ${\rm MeV}$): $g(^{40}{\rm Ca},n)=13(7)$,
$g(^{48}{\rm Ca},p)=17(6)$, $g(^{48}{\rm Ca},n)=-0.4(5)$ and
$g(^{56}{\rm Ni},n)=0.5(6)$. It follows that extruder-intruder
closures are non-existent, and harmonic oscillator closures are too
strong. A related problem is that the $0d_{5/2}$ hole orbits in
$^{47}$Ca and $^{47}$K are underbound by $\sim 4 \mev$ with respect to
their $sd$ counterparts. On the contrary the $1s_{1/2}-0d_{3/2}$
splittings are quite good: Although too large by some $1.5 \mev$ in
$^{39}$Ca, in $^{47}$Ca the splitting is drastically reduced, close to
what the data demand, and further reduced in $^{47}$K, now very close
to experiment.

Thus, we find that when the largest $j$ orbit in a major shell fills,
it binds itself and contributes to the binding of the largest $j$ orbits
in neighboring shells in a way that NN forces fail to reproduce. This
is the $\bm{j_{max}}$ anomaly. The necessary intra-shell self-binding
to cure it is now well understood in terms of a 3N
mechanism~\cite{Zuker:2003}. A mechanism to resolve the cross-shell
binding problem detected here remains to be found.  

In summary, we have shown that $\vlk$ and $G$ matrix elements are
quantitatively similar, but $\vlk$ as a free-space potential is far
easier to use in many-body calculations. As a consequence of the
similarity, it is possible to build on the successes of the $G$ matrix
approach, without the drawbacks due to the ill-defined starting
energies and other limitations on the SM by hard potentials.  $\vlk$
leads to matrix elements that are approximately cutoff and oscillator
frequency independent up to an overall scaling with the width of the
interaction. The scaling properties associated to the universal
behavior of $\vlk$ have been tested in \cspm spectra.  The soft nature
of $\vlk$ allows direct monopole estimates of binding energies and
\cspm spectra, and in an NN-only description, $\vlk$ reproduces many
features of the \cspm spectra. Moreover, our results suggest
that, apart from saturation, the main problem that demands a 3N
interaction is related to extruder-intruder shell formation. 

Finally, it is worth mentioning that ongoing NCSM
calculations~\cite{CSZ:2006} nicely confirm the validity of our $0\hw$
explorations. In particular, the saturation patterns for $A=6, 12, 16$
follow those of Fig.~\ref{fig:be} within an overall energy scaling 
and the spectra of $A=15, 17$ are the same within few hundred keV.
They provide some first suggestions on how nuclear
spectroscopy can help to constrain the 3N interactions that prove
crucial for a microscopic understanding of nuclear many-body systems.

We thank Hans Feldmeier and Dick Furnstahl for useful discussions. 
The work of AS is supported by the DOE Grant No. DEFG 0287ER40365 
and NSF Grant No. NSF--PHY 0244822.

\bibliography{pheno}
\end{document}